\documentstyle[prd,preprint,aps]{revtex}

\def\a{\alpha}

   \def\O{\Omega}

\def\slash#1{\,/\kern-7pt#1}
\def\rd{\partial}

\def\darr#1{\raise1.5ex\hbox{$\leftrightarrow$}\mkern-16.5mu #1}

\def\rds{/\kern-6pt\rd}

\newcommand{\be}{\begin{equation}}
\newcommand{\bea}{\begin{eqnarray}}
\newcommand{\ee}{\end{equation}}
\newcommand{\eea}{\end{eqnarray}}
\newcommand{\ba}[1]{\left(\begin{array}{#1}}
\newcommand{\ea}{\end{array}\right)}

\newcommand{\nn}{\nonumber}

\newcommand{\id}{\kern0.2em\rule{0.1mm}{0.71em}
                 \kern0.12em\rule{0.1mm}{0.71em}
                 \kern-0.27em\rule[0.68em]{0.27em}{0.1mm}
                 \kern-0.30em\rule{0.44em}{0.1mm}\rule{0.1em}{-1mm}}

\def\prd#1#2#3{Phys. Rev. D {\bf {#1}} ({#2}) {#3}}

\def\plb#1#2#3{Phys. Lett. {\bf B{#1}} ({#2}) {#3}}
\def\npb#1#2#3{Nucl. Phys. {\bf B{#1}} ({#2}) {#3}}

\def\pr{\prime}

\begin{document}
\draft
\preprint{KIAS-P98002, hep-th/9802026}
\title{The Background Geometry of DLCQ Supergravity}
\author{Seungjoon Hyun%
        \footnote{email: sjhyun@nms.kyunghee.ac.kr}}
\address{Department of Physics and Research Institute for Basic
        Sciences \\
        Kyunghee University, Seoul 130-701, Korea \\
        and \\
        School of Physics, KIAS, Seoul 130-012, Korea}
\date{February, 1998}
\maketitle
%\vskip 3.0cm

\begin{abstract}
By following Seiberg's prescriptions on DLCQ of $M$ theory, we give
 the  background geometries of DLCQ supergravity associated with $N$
sector
of DLCQ of $M$ theory on $T^p$ with vanishingly small radii. Most of
these are the product of anti-de Sitter
spacetimes and spheres, which have been found as the spontaneous
compactifications of eleven dimensional supergravity long time ago and
also are
revisited recently by Maldacena by considering the near horizon geometry
of
various D-branes in appropriate limit. Those geometries are maximally
symmetric and have full 32 supersymmetries of eleven dimensional
supergravity, which agrees with the number of supersymmetries of DLCQ of
$M$ theory. This suggests that DLCQ of $M$ theory is
the $M$/string theory on these nontrivial background.
\end{abstract}
\newpage
\section{Introduction}

One of the exciting developments for the formulation of eleven dimensional
$M$ theory has started with the $M$(atrix) model given in \cite{bfss}.
It is the formulation of $M$ theory in infinite momentum frame (IMF).
The basic idea is that, in this infinitely boosted frame, the only relevant
degrees of  freedom are those
of $N$ D0 branes and thus the effective action is given by
those of $N$ D0 branes which is
the dimensional reduction of ten dimensional $U(N)$
super Yang-Mills theory down to (0+1)-dimensions.
Later Susskind has noted \cite{susskind} that, instead,
one can consider the $N$-sector of light-cone compactification, so-called, the
discrete light-cone quantization (DLCQ) of $M$ theory which would be
identical to the
IMF matrix model in the large $N$ limit. Recently, the clear statement on this
matrix model conjecture of $M$ theory was given by Seiberg \cite{seiberg} and
Sen\cite{sen}. See also \cite{banks,bs} for nice reviews on $M$(atrix)
theory and \cite{hp,bs1}.  Following them, the finite $N$ DLCQ
matrix model, which is compactified on light-like circle with radius $R$ and
Planck scale $M_P$, is equivalent to $\tilde M$ theory with Planck scale
$\tilde M_P$ compactified on a spatial circle of radius $\tilde R$ in the
limit $\tilde R \rightarrow 0$, in which it becomes, naturally,
 the theory of $D0$ branes.

Recently, asymptotically anti-de Sitter ($AdS$) black hole solutions of
superstring theory have got much attention. Historically, they have been
realized as  the solutions of supergravity via the spontaneous
compactification over $S^n$. In their new perspectives, first noted in
\cite{mblacka}, they are realized as the solutions of type II superstring
theories and are  connected with the usual
asymptotically flat black hole solutions by the $U$-duality transformations.
See also \cite{bps,ss}.
Since it involves, asymptotically, light-cone compactification (i.e.
periodic in light-cone coordinate), it would be natural to consider the
solutions as those of DLCQ supergravity. Indeed, this kind of non-flat
background geometries is the natural one to appear as one considers the
supergravity limit of DLCQ of $M$ theory \cite{bbpt,mblackb}. By following
\cite{seiberg}, it has  been shown explicitly in
\cite{mblackb} that if one takes DLCQ on the
background five dimensional black holes  with three charges, one gets
effectively two-dimensional black hole solutions .

In this paper we describe the background spacetimes of the DLCQ supergravity,
corresponding to the DLCQ of $M$ theory on
$T^p$ with vanishingly small radius. We follow closely the prescriptions given
in \cite{seiberg} and \cite{sen} and
realize a wide class of the product of anti-de Sitter spacetimes and spheres
as background geometries of DLCQ supergravity.  Some of these are those found
long time ago as the solution of spontaneous compactifications of eleven
dimensional supergravity. In his recent paper \cite{mald-a}, Maldacena shows
that the same kind of various AdS spacetimes of supergravity emerges from the
large $N$ limit of certain conformal field theories. Based on this,
supplemented with other arguments, he conjectured that compactifications of
M/string theory on various anti-de Sitter spaces are dual to various conformal
field theories. We get these solutions as the background geometry of
supergravity of $N$ sector of DLCQ $M$ theory.

In section two, we summarize
some basic prescription of DLCQ of $M$ theory given in \cite{seiberg,sen}.
In section three, we consider the background geometries of various
compactifications of DLCQ $M$ theory over $T^p$. We get the
background spacetimes which are asymptotically $AdS_4\times S^7$,
$AdS_5\times S^5$, $AdS_7\times S^4$
and $R^{1,6}\times S^3$ for $p=2,3,4,5$, respectively. In section four, we
draw some conclusions.

\section{Brief Reviews on DLCQ of $M$ theory}
By following \cite{seiberg}, we consider the compactification of $M$ theory
on a light-like circle as a limit of a compactification on a small spatial
circle of radius $R_s$, boosted by a large amount.
As it has light-like compactification with the radius $R$, we can consider
$x^+$ as a genuine time coordinate.
After the
compactification along $x^-$, it becomes the ten-dimensional theory
with hidden eleventh dimension with radius $R$.
The prescription given in \cite{seiberg} and \cite{sen} is that this DLCQ
of $M$ theory is equivalent to another
$M$ theory, refered to as $\tilde M$ theory with different Planck scale
$\tilde M_P$ compactified on a small spatial circle of radius $\tilde R$
with the identifications,
\be
M_PR_i = \tilde M_P \tilde R_i,
\label{rel1}
\ee
 for the nine dimensional transverse space
part $x^i$
and
\be
\tilde R \tilde M_P^2 = R M_P^2.
\label{rel2}
\ee
These are finite in the limit $\tilde R\rightarrow 0$.
In this limit we have the string theory with string coupling and string
scale
\bea
&\tilde g_s=(\tilde R \tilde M_P)^{3\over 2} \sim \tilde R^{3 \over 4}
\rightarrow 0 \nn \\
&\tilde M_s^2=\tilde R \tilde M_P^3\sim\tilde R^{-{1 \over 2}}
\rightarrow \infty,
\label{rel3}
\eea
which becomes
the theory of $N$ D0-branes, living in a small transverse space of
characteristic size $\tilde R_i
\sim \tilde R ^{1 \over 2} \rightarrow 0$.

For a background with compactification on $T^p$
one uses T duality so that it  becomes the theory of
$N$ D$p$-branes on a torus with
radii
\bea
\tilde \Sigma_i = {1 \over \tilde R_i \tilde M_s^2} ={1 \over R_i
R M_P^3},
\label{rel4}
\eea
which are finite under the limit $\tilde R \rightarrow 0$.  The string
coupling after this T duality transformation becomes
\bea
\tilde g_s^\prime = \tilde M_s^{p-3} (RM_P^2)^3\prod \Sigma_i \sim
\tilde R^{-(p-3)/4}.
\label{rel5}
\eea
The low energy effective theory for $p\le 3$ is described by
$(p+1)$-dimensional SYM  with gauge coupling
\bea
 g_{YM}^2 = {\tilde g_s^\prime \over \tilde M_s^{p-3}} =
(R M_P^2)^3 \prod \Sigma_i,
\eea
which is finite under the limit $\tilde R \rightarrow 0$.
In the following section we use these to find the background geometry of
DLCQ supergravity for $M$ theory on $T^p$. As it turns out, it is not
the Minkowski spacetime, yet it has the full
supersymmetry of eleven dimensional supergravity.

\section{DLCQ of M theory on $T^p$}
We consider the metric of D0 branes on $T^p$ in
the infinite momentum frame. It is the straightforward generalization of
D0 branes on noncompact space given in  \cite{mblackb}.
The D0 brane configurations of the type IIA superstring theory on $T^p$
is given by \cite{hs}
\bea
ds^2 &=& - \frac{1}{\sqrt{f_0}} dt^2
+ \sqrt{f_0}(dx_1^2 + \  \cdots \  + dx_p^2  )
+ \sqrt{f_0}(dx_{p+1}^2 + \  \cdots \  + dx_9^2  ),\nn \\
e^{-2 \phi} &=& g_s^{-2}f_0^{(p-3)/2},
\label{d0}
\eea
where $g_s\equiv e^{\phi_\infty}$ is the string coupling constant,
$x^i$, $i=1\cdots p$, are internal $T^p$ coordinates with radii $R_i$
and a $(9-p)$-dimensional harmonic function
$$ f_0 = 1 +  (\frac{r_0}{r})^{7-p}, $$
 with the charge $r_0^{7-p}= \frac{Ql^9_p}{R_s^2R_1\cdots R_p}$
and $r^2 = x_{p+1}^2 + \  \cdots \ + x_9^2$.
The dimensionless number $Q$ is proportional to $N$, the number of
D-particles.

It has been shown in \cite{mblackb} that the field configurations in DLCQ
is given by
\bea
ds^2 &=& - \frac{1}{\sqrt{f}} dt^2
+ \sqrt{f}(dx_1^2 + \  \cdots \  + dx_p^2  )
+ \sqrt{f}(dx_{p+1}^2 + \  \cdots \  + dx_9^2  ),\nn \\
e^{-2 \phi} &=& g_s^{-2}f^{(p-3)/2},
\label{d0a}
\eea
with the new harmonic function $f$ given by
\be
f = \frac{Ql^9_p}{R^2R_1\cdots R_pr^{7-p}},
\label{harma}
\ee
where $R$ is the compactification radius over the light-cone circle $x^-$.
In the generic cases, when the locations of D-particles are
arbitrary, we get (\ref{d0a}) with
\be
f = \sum_a\frac{Q_a l^9_p}{R^2R_1\cdots R_p|\vec{r}-\vec{x_a}|^{7-p}},
\label{harma1}
\ee
where $Q=\sum_aQ_a$ and $r=0$ is the center of
mass of D-particles.

Since we are taking the light-cone coordinate $x^-$ as eleventh compactified
direction,
the string coupling and the string scale are given by
\bea
& g_s=( R  M_P)^{3\over 2} \nn \\
&M_s^2= R  M_P^3.
\label{rel11}
\eea

The prescriptions on the DLCQ of M theory
given in the  last section suggest that
this background geometry is connected  to the
background geometry of ${\tilde M}$ theory with the following
identifications for the metric,
\be
\frac{1}{\tilde \a^\prime}d{\tilde s}^2= \frac{1}{\a^\prime}ds^2,
\label{mtransa}
\ee
and the string coupling,
\be
\frac{\tilde g_s}{g_s}=
(\frac{\tilde \a^\prime}{\a^\prime})^\frac{3}{2}.
\label{mtransb}
\ee
In terms of the coordinates of ${\tilde M}$ theory, $\tilde x^m$,
(\ref{d0a}) with the harmonic function (\ref{harma}) become
\bea
d\tilde s^2 &=& - \frac{1}{\sqrt{\tilde f}} dt^2
+ \sqrt{\tilde f}(d\tilde x_1^2 + \  \cdots \  + d\tilde x_p^2  )
+ \sqrt{\tilde f}(d\tilde x_{p+1}^2 + \  \cdots \  + d\tilde x_9^2  ),\nn
\\ e^{-2\tilde \phi} &=& \tilde g_s^{-2}\tilde f^{-3/2},
\label{d0b}
\eea
where
\bea
\tilde f=
\frac{Q\tilde l^9_p}{\tilde R^2\tilde R_1 \cdots \tilde R_p
\tilde r^{7-p}}= (\frac{\tilde M_P}{M_P})^2 f,
\label{harma2}
\eea
and
$\tilde R_i=\frac{M_P}{\tilde M_P}R_i$ are radii of
$\tilde x^i$.
This, or more general cases with
\bea
\tilde f=
\sum_a\frac{Q_a\tilde l^9_p}{\tilde R^2\tilde R_1 \cdots \tilde R_p
|\tilde{\vec r}-\tilde{\vec x_a}|^{7-p}},
\label{harma3}
\eea
is indeed the supergravity solutions of
the $\tilde M$ theory with the same configurations
in the limit (\ref{rel3}), which
implies the validity of the prescription described in section 2
when applied to the background geometry of
supergravity associated with the DLCQ of $M$ theory.
This, in turn, tells us that in the DLCQ of $M$ theory we are considering
the $N$-sector of $D0$ branes whose background geometry is those given in
(\ref{d0a}) with the harmonic function $f$ given by (\ref{harma}) or
(\ref{harma1}).
Note that in order to deal with the light-cone quantization of field theories
in the flat Minkowski spacetime, we need
to introduce fictitious cutoff parameter to turn the light-like coordinate
into the space-like one \cite{hp,bs1}.
In this case, on the contrary,  $x^-$ can be considered as true eleventh
coordinate with the compactification
radius $R$ as background D0-particles provide well-defined metric
in which $x^-$ play the role as space-like coordinate.
In the followings we perform T-dualities along $T^p$, in the $\tilde M$ theory
framework,  and map to original DLCQ $M$ theory to get the
background geometries associated with the supergravity solutions of DLCQ
of $M$ theory on $T^p$. This turns out to be equivalent to taking T-dualities
directly on the solution (\ref{d0a}).

\subsection{DLCQ of $M$ theory on $T^3$ and $AdS_5\times S^5$}
Let us first consider the $N$ sector of DLCQ M theory on $T^3$.
As described in section 2,
after performing
T-dualities along $\tilde x^i$ directions on the metric (\ref{d0b}) of $\tilde M$
theory, which turns the configurations into those
of D3 branes,  the background geometry, with the harmonic function
(\ref{harma2}) becomes
\bea
d{\tilde s}^2 = {\tilde \a^\prime}\biggl[
\frac{U^2}{\sqrt{Q{\tilde g_s^\prime}}}(-d\tilde t^2+d\tilde x_1^2 + d\tilde x_2^2+  d\tilde x_3^2)
+ \sqrt{Q\tilde g_s^\prime}(\frac{dU^2}{U^2} + d\O_5^2 )\biggr],
\label{ads5}
\eea
where $U$ is the radial coordinate, which is well-defined in the limit
$\tilde R \rightarrow 0$
\be
U\equiv \frac{\tilde r}{\tilde \a^\prime}=(RM_P^3)r.
\ee
By transforming back to the original DLCQ of $M$ theory using
(\ref{mtransa}) and
$${\tilde g_s}^\prime=g_s^\prime= (RM_P^2)^3 \prod_i^3 \tilde\Sigma_i$$
we end up with the configurations
\bea
ds^2 =
\frac{r^2}{\sqrt{Q g_s^\prime}\a^\prime}(-dt^2+dx_1^2 + dx_2^2+  dx_3^2)
+ \sqrt{Q g_s^\prime}\a^\prime(\frac{dr^2}{r^2} + d\O_5^2 ).
\label{ads5a}
\eea
As mentioned earlier, this can be derived by performing T-dualities directly
on the configurations (\ref{d0a}).
Generically we use (\ref{harma1}) and get
the family of background geometries, which are asymptotically
$AdS_5\times S^5$, with the metric
\bea
ds^2 =  \frac{1}{\sqrt{f}}(- dt^2
+ dx_1^2 + dx_2^2  + dx_3^2  )
+ \sqrt{f}(dx_4^2 + \  \cdots \  + dx_9^2  ),
\label{d4a}
\eea
where
$$
f = \sum_a
\frac{Q_a g_s^\prime \a^{\prime 2}}{|\vec{r}-\vec{x_a}|^4},
$$
and $\vec x_a$ are moduli of DLCQ of $M$ theory on $T^3$.
Hence  the $N$-sector of
DLCQ $M$ theory on $T^3$, which is described by the fluctuations on the
$N$ D3 brane background and thus N=4 four-dimensional Yang-Mills theory,
would be related to the type IIB supergravity/superstring on the
asymptotically $AdS_5\times S^5$ background (\ref{d4a}).
The radii for $x^i$, $i=1,2,3$ are given by
$$
\Sigma_i = \frac{\a^\prime}{R_i},
$$
and thus $x^i$ become noncompact coordinates if the original tori
of DLCQ $M$ theory
shrink to zero size.
Then the background geometry is given by
asymptotically $AdS_5\times S^5$,
with no identification in spatial coordinates of $ AdS $
space.
In this case, the number of geometrical Killing spinors becomes
32\cite{kk,hls}\footnote{For compact
spatial coordinates, only half of supersymmetries survive\cite{pope}.},
thus may  be treated as true vacua of
the DLCQ of $M$ theory.

\subsection{DLCQ of $M$ theory on $T^4$ and $AdS_7\times S^4$}
Let us consider another example:  $N$ sector of DLCQ $M$ theory
on $T^4$. In this case,
as can be seen from (\ref{rel5}) with $p=4$, the
string coupling divergies, therefore we need to go  the strong
coupling limit of $\tilde M$ theory, which is an eleven dimensional
theory. D4 branes become M5 branes wrapping the eleventh dimension
whose radius is now finite and is given by
\bea
\tilde\Sigma_{11} = \frac{\tilde g_s^\prime}{\tilde M_s} =(R M_P^2)^3 \prod
\tilde\Sigma_i.
\label{rel6}
\eea
Therefore $N$ sector of DLCQ $M$ theory on $T^4$ becomes the theory of $N$ M5
branes in eleven dimensions
\cite{witten,strominger,rozali,brs,abkss,witten2,br97,ckv,gs,abs}.
The eleven dimensional Planck scale of $\tilde M$ theory becomes
\bea \tilde
M_P^\prime =\frac{\tilde M_s}{(\tilde g_s^\prime)^{\frac{1}{3}}}  =(\tilde
R^2\tilde R_1\cdots \tilde R_4\tilde M_P^9)^{\frac{1}{3}}\sim  \tilde
R^{-\frac{1}{6}} \longrightarrow \infty, \label{rel7}
\eea
and transverse radial coordinate $\tilde r$ scales by (\ref{rel1}), i.e.
$
\tilde r \sim \tilde R^{\frac{1}{2}},
$
thus well-defined radial coordinate in the limit, $\tilde R \rightarrow 0$,
can be written as
\be
U^2\equiv \tilde r(\tilde M_P^\prime)^3 = r (M_P^\prime)^3.
\ee
After taking T-dualities in $\tilde M$ theory, we
get the background geometries of the system of D4 branes
\bea
d\tilde s^2 &=&  \frac{1}{\sqrt{\tilde f}}(-dt^2
+ dx_1^2 + \  \cdots \  + dx_4^2  )
+ \sqrt{\tilde f}(d\tilde x_5^2 + \  \cdots \  + d\tilde x_9^2  ),\nn
\\ e^{-2\tilde \phi} &=& \tilde g_s^{-2}\tilde f^{1/2},
\label{d4b}
\eea
where
$$\tilde f = \frac{Q}{(\tilde M_P^\prime \tilde r)^3}
=Q(\frac{\tilde M_P^\prime}{U})^6.
$$
As mentioned earlier, the eleventh dimension has
finite radius (it is very large relative to the Planck scale
$\Sigma_{11}\tilde M_P^\prime \sim \tilde R^{-\frac{1}{6}}\rightarrow
\infty$), therefore it is more appropriate to consider the full eleven
dimensional background geometry of $\tilde M$ theory.
By combining boosting and
rescaling back
\be
(M_P^\prime)^2 ds^2=(\tilde M_P^\prime)^2 d{\tilde s}^2,
\label{mtransc}
\ee
we end up with the background geometry of the
DLCQ of $M$ theory on $T^4$ which is given by
\bea
d s^2_{11} = (M_P^\prime)^{-2}\biggl[
\frac{U^2}{Q^{1/3}}(-dt^2+dx_1^2 + \cdots  + dx_4^2+dx_{11}^2)
+Q^{\frac{2}{3}}(\frac{4dU^2}{U^2} + d\O_4^2)\biggr].
\label{m5b}
\eea
Again, in the case when the moduli $\vec x_a$ are non-vanishing,
we use (\ref{harma1}) and get the background geometries which
reduce to (\ref{m5b}) in the asymptotic region $r\rightarrow
\infty$.
As before if one starts with vanishingly small radii of $T^4$ tori, the dual
tori as well as the eleventh dimension become non-compact and
the metric describes $AdS_7\times S^4$. Therefore
$N$-sector of DLCQ of $M$ theory on vanishingly small $T^4$ describes
supergravity/$M$ theory on asymptotically $AdS_7\times S^4$.

\subsection{DLCQ of $M$ theory on $T^2$ and $AdS_4\times S^7$}
One can similarly work out for $T^2$ compactifications of DLCQ $M$ theory.
In this case, after T-dualities on $T^2$, from (\ref{rel5}), the eleventh
dimension in $\tilde M$ theory has vanishingly small radius
\be
\Sigma_{11}=(\tilde R_1\tilde R_2\tilde M_P^3)^{-1}\sim
\tilde R^\frac{1}{2}.
\label{rel8}
\ee
At first sight, this seems to imply that it is the theory of $N$ D2 branes in
ten dimensions. However one should note that the scale of eleventh dimension,
(\ref{rel8}), is the same order of the characteristic length scale of
transverse dimensions (\ref{rel1}), which implies underlying $SO(8)$ structure
of the theory. By going back to the DLCQ of $M$-theory on $T^2$ with
vanishingly small radii $R_i\rightarrow 0$, one can easily see that the
radius of eleventh dimension as well as
the radii
of dual tori ${\hat T}^2$ become non-compact, indicating the geometry is
the eleven dimensional one.
In this eleven dimensional geometry, the eleventh direction becomes the
transverse
one and therefore the seven-dimensional harmonic function becomes
the eight-dimensional one and thus can be written as
$$ f=\frac{Q l^9_p}{ R^2 R_1  R_2 r^5}
=\frac{Q l_p^{\prime 6}}{\Sigma_{11} r^5}\longrightarrow
\frac{Q l_p^{\prime 6}}{ r^6}.
$$
The background metric of DLCQ supergravity for the system of $N$ M2 branes
becomes those of $AdS_4\times S^7$,
\bea
d s^2_{11} = ( M_P^\prime)^{-2}\biggl[
\frac{U^2}{Q^{2/3}}(-dt^2+dx_1^2 + dx_2^2)
+Q^{\frac{1}{3}}(\frac{dU^2}{4U^2} + d\O_7^2)\biggr],
\label{m2}
\eea
where $U=r^2(M_P^\prime)^3$.
As before, if we start with (\ref{harma1}) considering the cases
nonvanishing $\vec x_a$, we end up with the background geometries
which are asymptotically $AdS_4\times S^7$.

\subsection{DLCQ of $M$ theory on $T^5$ and $R^{1,6}\times S^3$}
Finally we consider DLCQ $M$ theory on $T^5$ \cite{brs,seiberg1}.
In the $\tilde M$ theory,
 after T-duality along $T^5$, one can see from (\ref{rel5}) that it
becomes the strong coupling
limit of D5 branes in type IIB theory.
Therefore we need to perform the S-duality and then it becomes the theory of
$N$  NS5 branes in type IIB theory with vanishingly small string coupling,
$$
\bar g_s=(\tilde g_s^\pr)^{-1}\sim \tilde R^{\frac{1}{2}},
$$
and finite string scale,
$$
 M_s^2=\frac{\tilde M_s^2}{\tilde g_s^\pr}=R^2(R_1\cdots R_5)M_P^9.
$$
The appropriate background geometry can be found accordingly, by using
(\ref{mtransa}), and in the limit, $vol(T^5)\rightarrow 0$ of original $T^5$
torus, it is given by the product of Minkowskian
spacetimes $R^{1,6}$ and $S^3$,
\bea
ds^2_{10} &=& - dt^2 +dx_1^2 + \  \cdots \  + dx_5^2
+ \frac{Q}{M_s^2}d\rho^2 + \frac{Q}{M_s^2}d\Omega_3^2,
\label{d5b}
\eea
where the well-defined radial coordinate
$U= \frac{\tilde r}{\bar g_s}= \frac{r}{ g_s}$ is replaced by
$\rho=\log (M_s U)$. As shown in \cite{mblackb}, if we have additional
branes, the geometry describes effective two-dimensional CGHS type black
holes \cite{cghs}. For example, consider the same compactification with
extra D4 branes.
 Along
the same line of arguments with S-duality and T-dualities on the
transverse directions, we end up with
the system of NS5 branes with extra fundamental strings, whose geometry is
given by  the metric of the form,
\bea
ds^2_{10} &=& \frac{1}{f_1}(- dt^2+ dx_1^2)
+ dx_2^2 + \  \cdots \  + dx_5^2
+Q\a^\pr\frac{dU^2}{U^2} + Q\a^\pr d\Omega_3^2,
\label{d5c}
\eea
where
$$
f_1=1+\frac{Q_4\Sigma_5(RM_P^2)^3\a^{\pr 2}}{U^2}.
$$
Here $Q_4$ is proportional to the number of D4 branes which turns into
fundamental strings after the chain of S- and T-dualities.

\section{Conclusions}
In this paper we have described the background geometry of DLCQ
supergravity associated with DLCQ $M$ theory. In contrast to the usual
situation, like in ordinary string theory, in which we consider the flat
Minkowskian spacetime as a true vacuum with full supersymmetry and consider
various excitations on those backgrounds, the N sector of DLCQ $M$ theory
should be interpreted as having N D0 branes (for non-compact spacetimes
and Dp-branes on $T^p$) as background with full supersymmetries. Therefore
various spacetimes found above should be considered as the background
geometries on which the theory should be defined.
Our results remain true for any $N$, though we generally need $N$
to be large in order to be in supergravity regime.
In the large $N$ limit, $M$(atrix) theory
conjecture tells us that it is equivalent to the $M$ theory with non-compact
eleventh dimension. Our results, which also holds in the large $N$ limit,
suggest that the $N$-sector of DLCQ of $M$ theory and its large $N$ limit
should be considered as  $M$/string theory defined on these
non-trivial, typically asymptotically $(AdS)_p\times S^q$, background.
The  derivation of \cite{seiberg,sen} is independent of the number of
supersymmetries.
Therefore all the discussions in this paper can be applied for
the non-supersymmetric cases as well.
For example one may begin with D0-anti-D0 bound
state in DLCQ on $T^p$. Then by following the above procedures,
one ends up with non-extremal anti-de Sitter
black hole solutions.

\acknowledgements{
I have benefited from comments by Jin-Ho Cho, Youngjai Kiem, Sangmin Lee and
Jae-Suk Park.
I also would like to thank Juan Maldacena who pointed out an error in previous
version. This work was supported in part by Korea Research
Foundation. }

%---------------------------------------------------------

%%%%%%%%%%%%%%%%%%%%%%%%%%%%%%%%%%%%%%%%%%%%%%%%%%%%%%%%%%%%%%%%%%%%%%%%%%
\end{document}